\documentclass[]{spie}

\usepackage{amsmath,amsfonts,amssymb}
\usepackage{graphicx}
\usepackage[colorlinks=true, allcolors=blue]{hyperref}
\usepackage{siunitx}

\title{Automated,
deep reactive ion etching free 
fiber coupling to nanophotonic devices}
\author[a,*]{Fabian Flassig}
\author[b,*]{Rasmus Flaschmann}
\author[a]{Thomas Kainz}
\author[a]{Sven Ernst}
\author[a]{Stefan Strohauer}
\author[b]{Christian Schmid}
\author[b]{Lucio Zugliani}
\author[b]{Kai M\"uller}
\author[a]{Jonathan J. Finley}

\affil[a]{Walter Schottky Institute and Physics Department, Technical University of Munich, 85748 Garching, Germany}
\affil[b]{Walter Schottky Institute and Department for Electrical and Computer Engineering, Technical University of Munich, 85748 Garching, Germany}

\authorinfo{*, Fabian Flassig and Rasmus Flaschmann contributed equally to this work. \\Correspondence to fabian.flassig@wsi.tum.de or rasmus.flaschmann@wsi.tum.de}

\begin{document}
\maketitle
\begin{abstract}
Rapid development in integrated optoelectronic devices and quantum photonic architectures creates a need for optical fiber to chip coupling with low losses. 
Here we present a fast and generic approach that allows temperature stable self-aligning connections of nanophotonic devices to optical fibers.
We show that the attainable precision of our approach is equal to that of DRIE-process based couplings. Specifically, the initial alignment precision is $1.2\pm\SI{0.4}{\micro\meter}$, the average shift caused by mating $<\SI{0.5}{\micro\meter}$, which is in the order of the precision of the concentricity of the employed fiber, and the thermal cycling stability is $<\SI{0.2}{\micro\meter}$.
From these values the expected overall alignment offset is calculated as $1.4 \pm \SI{0.4}{\micro\meter}$.
These results show that our process offers an easy to implement, versatile, robust and DRIE-free method for coupling photonic devices to optical fibers.
It can be fully automated and is therefore scalable for coupling to novel devices for quantum photonic systems.
\end{abstract}


\section{Introduction}
Recent years have seen major advances in photon based quantum science and technologies\cite{Kim08} such as quantum simulation and computation\cite{Svo2016, Sin2016}, distributed quantum technologies such as quantum key distribution \cite{BB84, Tak2007, Shi2014} and state teleportation \cite{Pfa2014} or deep space optical communication \cite{Cal2019, Iva2020}. 
Hereby, scalable approaches require the linking of on-chip components such as single-photon emitters (e.g. quantum dots \cite{Fry2000}, NV-centers in diamond \cite{Kur2000} or 2D materials \cite{Tra2016}) or quantum detectors (e.g. single-photon avalanche photodiodes (SPADs) \cite{Nix1932}, transition edge sensors (TES) \cite{Ull2015} or superconducting nanowire single-photon detectors (SNSPDs)\cite{Gol01, Rei13,  Miller:11, Redaelli_2016}).
To interface these building blocks as well as interconnecting them, coupling to optical fibers is the most common approach.
Nonetheless, efficient chip to fiber interconnects with minimum losses remain a major challenge. 
For example, any losses at interfaces will reduce the attainable bit rates for single-photon QKD \cite{Lee2019, Kupko2020, kupko2021evaluating}. 
Thus, the photonic design has to be engineered to minimize interface reflections especially at semiconductor substrates that have high refractive indices.\cite{Strauf2010, Mantynen2019} 
Beyond that, mechanical and temperature stability of a fiber to chip connection have to be guaranteed to provide ruggedness against environmental influences even under extreme conditions e.g. below \SI{5}{\kelvin} \cite{Comyn2018} for applications such as SNSPDs. 
In general, there are two ways of coupling an optical fiber to photonic elements. Either via 1) physical contact, butt or evanescent coupling or 2) via free space coupling using lenses. 
For any application that requires temperature stability, method 1) evidently exhibits less drift due to its more compact design using fewer optical components causing lower losses at interfaces. 
On the downside, the interaction length of the light with the photonic elements is limited by the thickness of the latter, while the illuminated area is defined by the fiber mode field diameter (MFD).
While additional optical resonators \cite{Mik2013, Zha17} can boost the effective interaction length significantly, various fiber coupling approaches have been proposed such as single-mode optical fiber taper \cite{Lie2014, Bur2016}, single-sided conical tapered optical fibers \cite{Tie2015}, GRIN lenses\cite{Mik2010, Wan2011}, 3D-printed couplers \cite{Geh2019, Geh2019.2, Xux2021} or the deep reactive ion etching (DRIE, also known as Bosch) process \cite{patent:5501893, Miller:11, Redaelli_2016} requiring a special RIE system and process. 
Therefore, it is of the utmost interest to develop a fast, simple, scalable and reproducible fiber coupling mechanism.

\section{Coupling mechanism}
\begin{figure}[htb]
\centering\includegraphics[width=\textwidth]{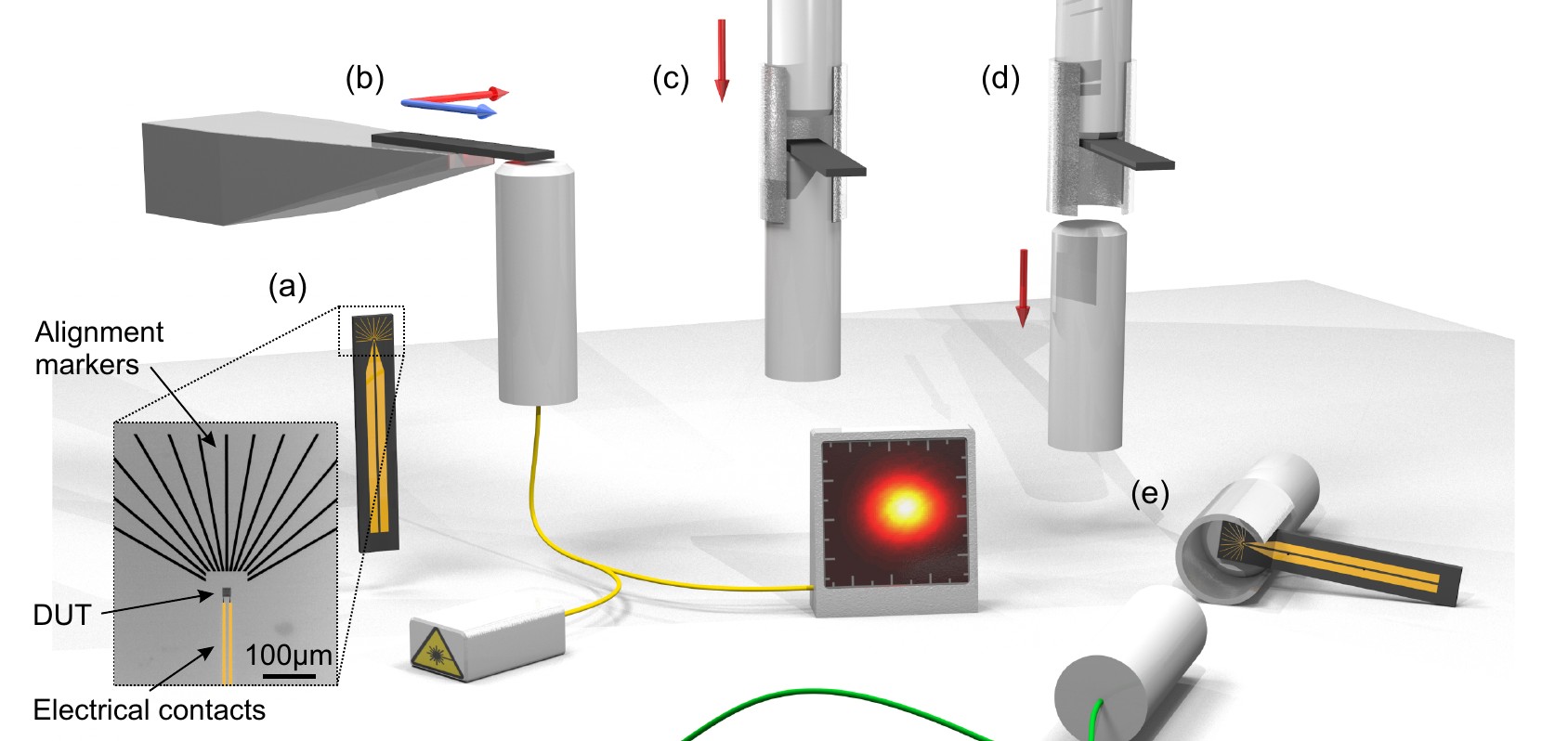}
\caption{Schematic illustration of fiber coupling mechanism. (a) Bare sample is cut/cleaved into a thin slice. Inset: Microscope image of sample with device under test (DUT), electrical contacts in gold and optical alignment markers in black. (b) Sample is mounted face down on translation stage and aligned with respect to a test fiber by recording the back-reflected light intensity in the fiber. (c) Guided via a mating sleeve, a blank ferrule is glued on the back side of the sample. (d) The alignment fiber is detached, (e) allowing to couple any identical fiber (green) to the sample.}
\label{fig:overview}
\end{figure}

The star-shape alignment fiber coupling mechanism (SAM) we present here involves the use of a star-shaped alignment structure, as depicted schematically in fig. \ref{fig:overview}.
Similar to other demonstrated mechanisms like self-aligned fiber coupling based on the Bosch process\cite{patent:5501893, Miller:11, Zad2021}, we make use of the well specified alignment properties of a zirconia mating sleeve. 
However, these mechanisms require the sample to be processed to fit exactly into the mating sleeve such that the alignment is purely delivered by the shape of the sample, the centricity of the optical component and the quality of the sidewalls. 
For this, a multi-step DRIE process is necessary to create the sample shape with the required precision. 
This highly elaborate process limits the choice of substrate materials to silicon and requires careful engineering. 
We eliminate this necessity and start in fig. \ref{fig:overview}(a) with a sample that has been cleaved to precisely fit into the gap of a mating sleeve ($\leq$\SI{500}{\micro\m}).
The inset shows a microscope image of a sample with a nanodevice in the center (here a SNSPD), electrical contacts in gold and star-shaped alignment markers in black.
These markers can either be fabricated in the same fabrication step as the nanodevice or after a post-fabrication selection that allows identification of the best performing nanodevice.
Note that if these star-shaped markers are fabricated in a second step, the alignment offset can accumulate during the overall alignment process.
The sample is subsequently mounted face down on a sub-micron precision three axis translation stage.
As shown in fig. \ref{fig:overview} (b) it is positioned above the end facet of a split fiber of the same type that will later be coupled to the sample.
This fiber is split into two ports, one is connected to a laser at a chosen alignment wavelength to illuminate the sample, the other is connected to a photodiode to record the light intensity reflected from the sample surface. 
The contrast in reflectivity between sample surface and alignment structure is utilized to position the fiber MFD relative to the sample. 
As depicted schematically in fig. \ref{fig:overview}(c), once alignment has been achieved a mating sleeve is attached to the fiber and a blank stabilization ferrule is glued onto the back side of the sample.
Once the glue has hardened, the split fiber and the actuator can be removed, leaving the sample glued to the blank ferrule as shown in fig. \ref{fig:overview}(d).
This device can then be installed in any experimental setup, e.g. a cryostat.
Here, the blank ferrule facilitates the alignment to the target fiber as depicted in fig. \ref{fig:overview}(e). The protruding section of the sample allows for electrically contacting the nanodevice.

Centering the fiber core relative to the device position is crucial for proper coupling.
As depicted schematically in fig. \ref{fig:overview}(b), we aim for a mechanical alignment process that scans the sample surface utilizing the fiber itself by sending light through the fiber and a fused fiber 1x2 beam splitter \cite{Thorlabs-Y-Splitter} and onto the sample surface.
Then, the light is reflected back through the beam splitter onto a photodiode allowing to measure the back-reflected intensity.
The Gaussian beam scanning the surface has a wavelength-dependent mode field diameter, where the intensity drops to $1/e^2 = 4 \sigma$.
Here, $\sigma$ corresponds to the standard deviation of a Gaussian beam normal distribution.
Due to the wavelength-dependence of the MFD we express all lengths in the following paragraphs in terms of multiples of $\sigma$, allowing for structures to be tailored to a specific wavelength. 
In the following, we consider circular DUTs with the area $A$ and the radius $r$. In this framework we define the mode coverage $\varphi$, i.e. the overlap integral between the DUT and the Gaussian mode, as
\begin{equation}
	\varphi(x_0) = \frac{1}{A} \iint_A H(x,y) \cdot \exp\left( - \frac{x^2}{2\sigma^2} - \frac{y^2}{2\sigma^2}\right)\, \mathrm{d}x \, \mathrm{d}y \text{ , where } H(x,y) = \begin{cases}
1 &\text{for } (x-x_0)^2 + y^2 \leq r^2\\
0 &\text{otherwise}
\end{cases}
	\label{eq:relative_absorption}
\end{equation}
is a 2D Heavyside function for the area integral over the Gaussian mode field intensity distribution. 
Moreover, $A = (x-x_0)^2 + y^2 \leq r^2$ is defined as a disk with $x_0$ as the offset from the maximum of the mode field.
To measure the alignment precision of our method, we fabricated circular DUTs with varying radii surrounded by the star-shaped alignment markers on a transparent substrate to allow for different reflective alignment procedures while providing optical feedback via a microscope.
The structures themselves are made of a \SI{50}{\nano\meter} thick gold layer evaporated on top of a \SI{10}{\nano\meter} titanium adhesion layer.
These structures are patterned using electron beam lithography and liftoff before metallization.
To conduct the alignment process, the fiber tip is brought within $\sim\SI{5}{\micro\meter}$ to the sample surface.
\begin{figure}[htb]
\centering\includegraphics[width=\textwidth]{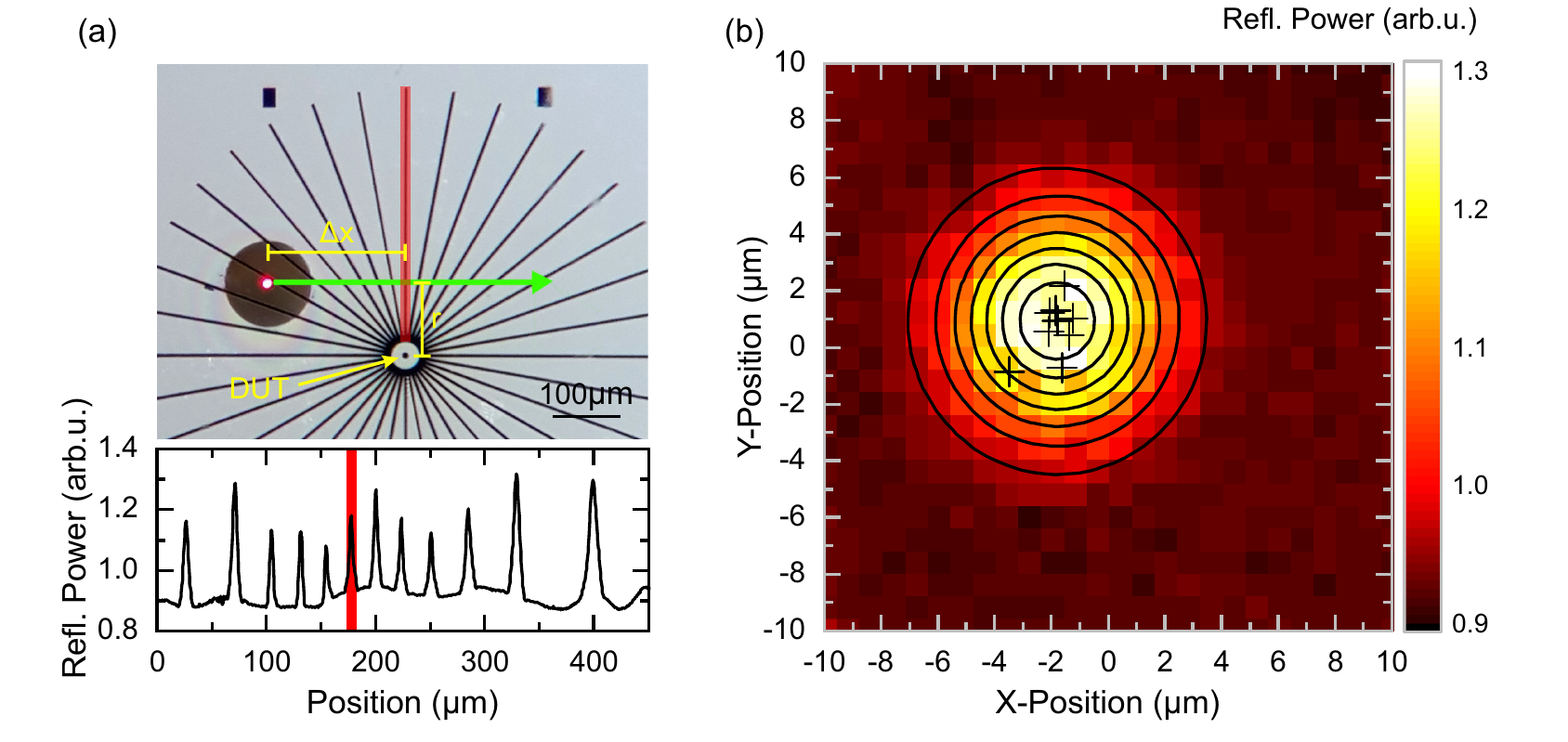}
\caption{(a) Top: Alignment process photographed for a transparent substrate; the fiber (cladding visible as dark grey circle, core illuminated with \SI{635}{\nano\m} laser diode) is linearly moved across a star-shaped alignment structure around the DUT, here a \SI{7}{\micro\meter} diameter gold disk. Bottom: intensity profile of back reflected light in optical fiber traversing the green arrow in the upper picture. The red lines in top and bottom mark the same feature. (b) To analyze placement precision, a 2D piezo-actuated scan is performed while the measuring the back reflected intensity. The black rings correspond to a 2D Gaussian fit of the reflection pattern of the target, the crosses mark the centers of fits of multiple runs.}
\label{fig:auto_mechanism2}
\end{figure}
Then, a 1D line scan is performed across the alignment structures along the direction of the green arrow as shown in fig. \ref{fig:auto_mechanism2} (a). 
The corresponding reflected light intensity is plotted in the bottom panel fig. \ref{fig:auto_mechanism2} (a) as a function of the fiber position.
Due to the interference between the optical fields propagating in the fiber and reflection from the sample surface, the baseline of the signal slightly fluctuates, while the peaks are readily distinguishable.
Hereby, the FWHM of these peaks varies from \SIrange{4}{8}{\micro\meter} depending on the angle between scan direction and alignment lines as well as the varying distance to the neighboring peaks. 
The position of each peak is given as the intersection of the line scan ($x_\mathrm{i}$) with the corresponding peak via
\begin{equation}
    x_\mathrm{i}=r\cdot \tan(\alpha_\mathrm{scan}\cdot n_\mathrm{i}-\alpha_0)+ \Delta x
    \label{eq:peak_pos}
\end{equation}
where $r$ is the distance of the scan line to the star center and $\Delta x$ the offset of the scan start point along the scan line from the central strip (highlighted in red in fig. \ref{fig:auto_mechanism2} (a)). 
Here, $\alpha_\mathrm{scan}$ is the angular step between two star-shaped alignment lines (i.e. $10^{\circ}$) and $\alpha_\mathrm{0}$ is the offset angle between the scan direction and sample orientation. 
To fit the peak positions with eq. (\ref{eq:peak_pos}), a Savatzky-Golay non-linear curve fit \cite{sg} was used to retrieve the offset of the fiber core from the center of the DUT. 
Note that the symmetry of a full star structure leads to an ambiguity in the orientation of the center. 
Therefore, we restrict the alignment structure to its upper half.
After driving to the calculated center position (cf. eq. (\ref{eq:peak_pos})), a local 2D piezo-actuated scan is performed to map out its position relative to the fiber. 
A typical map of the reflected intensity for an area scan over $20\times$\SI{20}{\micro\meter} and a gold disk diameter of \SI{7}{\micro\meter} is presented in fig. \ref{fig:auto_mechanism2}(b).
For repeated alignment procedures the relative central positions are marked by black crosses, showing a systematic offset of \SI{2}{\micro\meter} at an alignment precision of $1.2\pm\SI{0.4}{\micro\meter}$. 
The systematic offset has been apparent throughout all measurements and originates from a backlash in the motor stages that could also be compensated for in the algorithm. 
Overall, the combination of sub-\SI{2}{\micro\meter} precision, few prerequisites and high alignment speed makes this a highly versatile process for fiber-to-device alignment. 
The high-resolution piezo scanning option can further enhance the resolution to the sub-micron regime. 
The sample can then be temporally fixed to the alignment fiber using a droplet of soluble glue for enhanced stability before completing the steps depicted in fig. \ref{fig:overview}(c).

\section{Alignment Precision and stability}

Since the aim is to develop an automated fiber coupling procedure that is not only fast, simple and scalable but also reproducible, investigating the achievable alignment precision is of high interest.
In addition, for all cases in which light shall be collected from, and not emitted into  the fiber, the required size of the optical device (e.g. a grating coupler \cite{Pig2014} or SNSPD\cite{Gol01, Rei13,  Miller:11, Redaelli_2016}) is determined by both the MFD and the alignment precision. 
\begin{figure}[htb]
\centering\includegraphics[width=\textwidth]{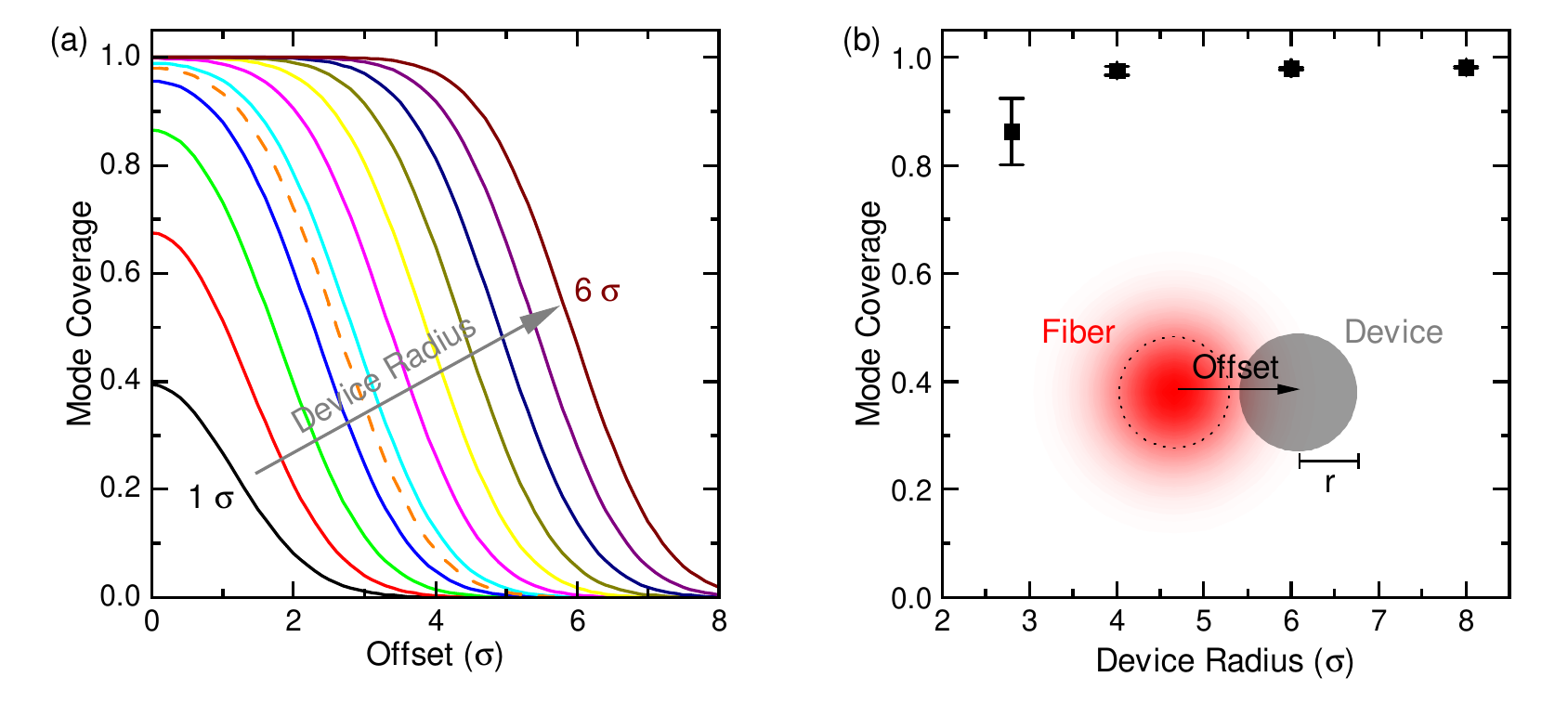}
\caption{(a) Simulated mode coverage as a function of fiber-to-detector offset for different detector radii normalized to the variance ($\sigma^2$) of the Gaussian beam profile ($d_{1/e^2} = 4\sigma$) for a 780HP fiber with $\sigma = \SI{1.25}{\micro\meter}$. The dashed orange line corresponds to a device radius of $2.8 \sigma$.
(b) Measured mode coverage as a function of device radius for samples coupled with the star-shape fiber coupling mechanism. The error bars represent the standard deviation for n = 5 alignment runs. Inset: schematic depiction of a misaligned circular structure (grey) and the fiber core (dashed circle) with its beam mode (red), as used both for the theoretical calculation in (a) and the measurement in (b).}
\label{fig:auto_mechanism_data_1}
\end{figure}

We start by presenting an analysis of the light absorption in the nanodevice depending on the device radius as well as a possible offset between fiber and device. This allows the optimization of the device size depending on the operating wavelength and alignment precision.
Fig. \ref{fig:auto_mechanism_data_1}(a) depicts the mode coverage by the DUT $\varphi$ (eq. (\ref{eq:relative_absorption})) as a function of the alignment offset $x_0$ for different radii from $1 \sigma$ (black) to $6 \sigma$ (brown), as schematically depicted in the inset in panel \ref{fig:auto_mechanism_data_1}(b). 
Considering a 780HP fiber with a MFD of \SI{5}{\micro\meter} at \SI{780}{\nano\meter}\cite{Fiber780HP}, this corresponds to device diameters ranging from \SIrange{2.5}{15}{\micro\meter}. 
For a perfect alignment with zero offset, the mode coverage by the DUT corresponds to the cumulative distribution function of a 2D normal distribution.
Thus, a device radius of at least $3 \sigma$ (turquoise line on fig. \ref{fig:auto_mechanism_data_1}(a)) is required to ideally reach a mode coverage above $99\%$.
Increasing the diameter of the optical device to a diameter of \SI{15}{\micro\meter} ($r = 6 \sigma$) allows for an offset of $\pm\SI{5}{\micro\meter}$ ($4 \sigma$) while still providing a mode coverage of $\varphi \geq 96\%$. 
The offset can stem from e.g. misalignment, thermal strain or vibrations of the cold stage of a pulse-tube cryocooler.
We experimentally measure the mode coverage for gold disks on a glass substrate in a transmission geometry. Hereby, a mode coverage of 1 corresponds to a complete blocking of the light from the fiber by the DUT and a mode coverage of 0 corresponds to the unhindered transmission of the light through the glass substrate when no DUT is present. For different DUT sizes the average mode coverage for n = 5 alignment runs is presented in fig. \ref{fig:auto_mechanism_data_1}(b). 
All sizes are given in multiples of $\sigma$ with $1 \sigma = \SI{1.25}{\micro\meter}$ for the 780HP fiber at \SI{780}{\nano\meter}.
Hereby, the  dashed orange line of fig. \ref{fig:auto_mechanism_data_1}(a) represents the theoretical curve for the first data point with $2.8 \sigma$. 
For larger structure sizes, the mode coverage reaches $98\%$, in agreement with the transmittance of a \SI{50}{\nano\meter} thick gold film. The onset of a misalignment for the first data point at \SI{7}{\micro\meter} disks can be calculated using eq. (\ref{eq:relative_absorption}) to be a statistical misalignment of $1.8\pm\SI{0.4}{\micro\meter}$ which is in agreement with the previously measured value of $1.2\pm\SI{0.4}{\micro\meter}$.

\begin{figure}[htb]
\centering\includegraphics[width=\textwidth]{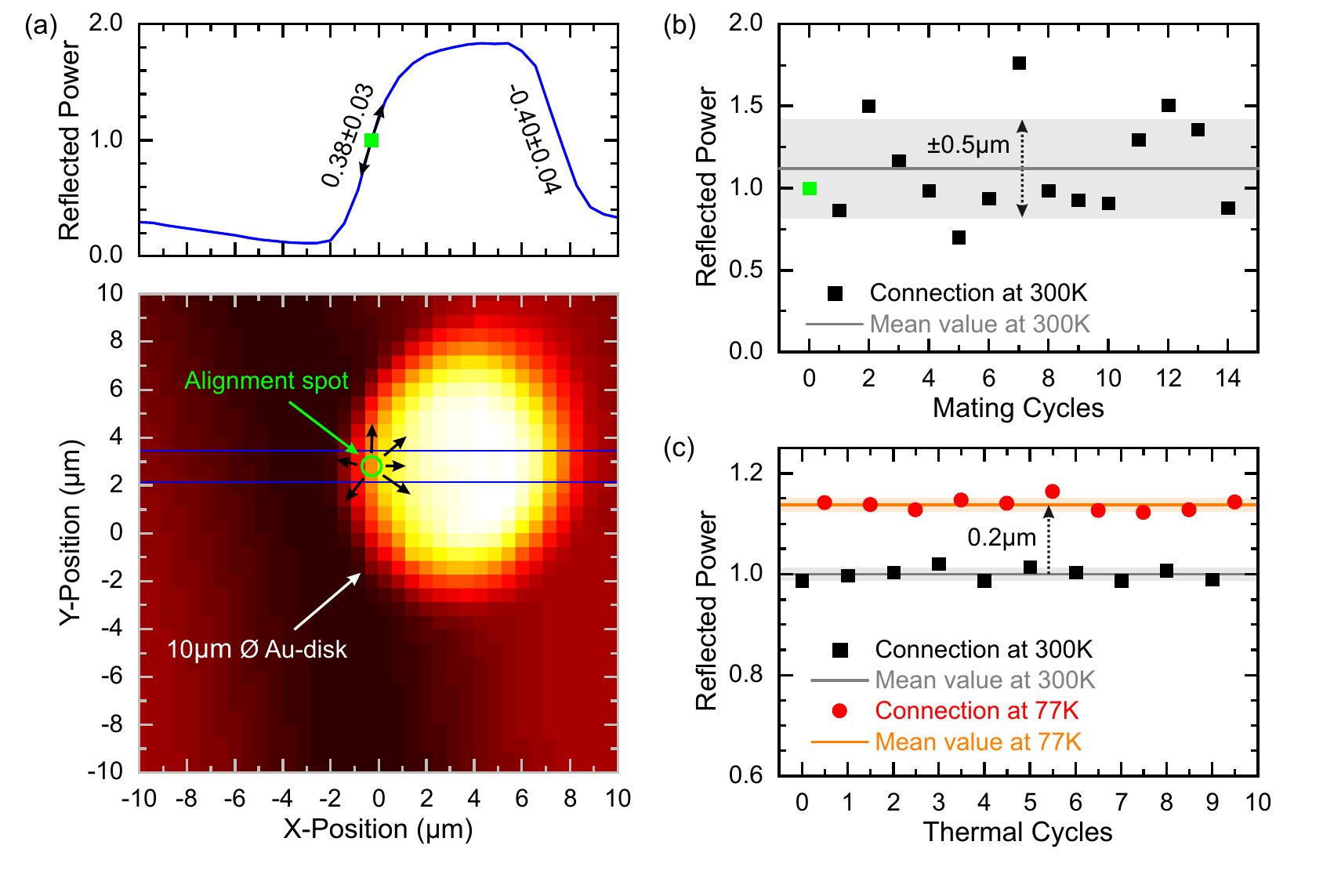}
\caption{(a) Bottom panel: 2D scan of a \SI{10}{\micro\meter} gold disk in reflective geometry. The blue lines indicate the cross section depicted in the upper panel. For stability measurements the fiber is aligned onto the edge of the disk and coupled at this position (green circle). Top panel: corresponding 1D cross section. Indicated is a linear fit through the sidewalls of the plateau on both sides with a slope of \SI{0.4}{\per\micro\meter}. (b) Reflected power normalized to its mean value (dark grey line) for multiple mating cycles at the same position ((a), green circle). The measurement is performed at room temperature (RT) by unplugging and reconnecting an optical fiber in random axial orientations. The standard deviation is depicted as the light grey shaded area together with its estimated corresponding shift $\pm\SI{0.5}{\micro\meter}$. (c) Reflected power normalized to the mean value at RT for multiple thermal cycles. The sample and connector are dipped into liquid Nitrogen and reheated to RT. In black/grey the reflected power at RT, its mean and standard deviation is shown. In red/orange the same is shown at \SI{77}{\kelvin}.}
\label{fig:alignment_stability_v3}
\end{figure}

To allow for a quantitative analysis of the accuracy, stability and reproducibility of this alignment method, a sample with a \SI{10}{\micro\meter} diameter gold disk ($ 4 \sigma$, cf. yellow line in fig. \ref{fig:auto_mechanism_data_1}(a)) was prepared.
In fig. \ref{fig:alignment_stability_v3}(a) we present data of a 2D piezo-actuated scan (bottom panel) with a cross section along the x-axis through the center of the disk (blue, top panel).
The cross section reveals a high reflected power with a plateau and two transition areas where the gold disk was illuminated.
The plateau is visible at the center of the disk with a size around $ 4 \sigma$ as expected when comparing with the yellow line in fig. \ref{fig:auto_mechanism_data_1}(a).
It allows for alignment tolerances as long as the MFD stays entirely within the plateau, as desired for coupling a fiber to a DUT.
The broad dark red/red stripes in the background of the 2D map as well as the slope within the plateau and the low reflected power regime of the blue cross section originate from interference patterns caused by a relative tilt of the sample.
However, to quantitatively test the reproducibility of the alignment, the plateau behavior in reflected intensity hinders the detection of small alignment shifts.\footnote{The size of this plateau can be reduced by smaller disks. However, the gradient of the mode coverage at $0\sigma$ offset is significantly lower than for a finite offset (cf. green line with $2\sigma$ offset in fig. \ref{fig:auto_mechanism_data_1}(a)). Thus, it prevents a sufficient detection of small fiber to DUT shifts $\ll 1 \sigma$.} 
Instead we align the fiber on the edge of the disk (highlighted by the green circle in the bottom panel of fig. \ref{fig:alignment_stability_v3}(a)) where the intensity shows the highest gradient.
Here, any relative shift between fiber and sample leads to a measurable change in reflected intensity. This alignment point is also indicated in green in the top panel of fig. \ref{fig:alignment_stability_v3}(a) together with the slope of the plateau sidewalls. 
We calculate the average slope for both x and y direction via a linear fit and determine the relative change in reflected intensity
\begin{equation}
\label{eq:dP_rel}
\biggl| \frac{ dP_\mathrm{rel}}{dr}\biggl| = \SI{0.37}{\per \micro \meter}
\end{equation}
when shifting the fiber along the gradient. This approach is of course only valid for small shifts around the marked position assuming a constant gradient pointing towards the structure center.
Thus, we can convert relative changes in the reflected power to an alignment shift $\overline{\Delta r}$.
However, eq. (\ref{eq:dP_rel})  only considers shifts along the gradient. 
Any shift perpendicular to the gradient will in a first-order approximation result in a constant intensity.
To compensate for this effect, we will randomize the orientation in which we couple the fiber relative to the sample, leading to randomly distributed directions of the alignment shifts.
The amplitude of these shifts hereby is independent of the direction. Therefore, both directions, along and perpendicular to the gradient, are independent random variables with the same mean value and standard deviation, giving a mean total shift of 
\begin{equation}
\label{eq:dP_tot}
\overline{\Delta r} = \sqrt{\overline{\Delta x}^2 + \overline{\Delta y}^2} = \sqrt{2} \cdot \overline{\Delta x}.
\end{equation}

As a next step we make use of this quantitative analysis model to investigate the reproducibility of the alignment for multiple mating cycles. 
Fig. \ref{fig:alignment_stability_v3}(b) shows the reflected power when a fiber is connected to the device repeatedly (n=15) in random orientations. The reflected power is normalized to the initial reflected power after the alignment process (green point). The mean value of the reflected power is indicated by the dark grey line and the standard deviation by the grey shaded area.
The intensity shows a significant spreading ($P_{ref} = 1.13 \pm 0.29$) as expected, since we are performing the analysis at the edge of the test structure at the alignment spot.
Even though we cannot deduce the absolute shift of a single data point, we can determine the statistical shift following eqs. \ref{eq:dP_rel} and \ref{eq:dP_tot} to be $\overline{\Delta r} = 0.46 \pm \SI{0.12}{\micro \meter}$.
This value agrees well with the specification of the fiber core/cladding concentricity of $\leq \SI{0.5}{\micro\meter}$.
We conclude that the performance of the mating sleeve is not influenced significantly by the presence of a sample in between both ferrules after multiple mating cycles. Most importantly, no significant systematic shift in the alignment is apparent after these mating cycles, indicating a very stable alignment.

The thermal stability of the connection was tested by cycling the temperature between room temperature (RT) and $\SI{77}{\kelvin}$ as shown in fig. \ref{fig:alignment_stability_v3}(c).
An off-the-shelf fast curing epoxy adhesive was used to glue the sample.
Dipping the connector into liquid nitrogen allows for a rapid cool down and strong thermal shock whilst simultaneously allowing monitoring of the whole process for a possible structural failure.
The same sample with the fiber aligned to the disk edge was used. This time, the fiber remained connected to the sample in the same orientation during the whole measurement.
Fig. \ref{fig:alignment_stability_v3}(c) shows the reflected power for ten thermal cycles from RT (black squares) to $\SI{77}{\kelvin}$ (red dots) normalized to the mean value at RT.
The corresponding mean values and standard deviations of both data sets are indicated by grey and red lines and shaded areas, respectively.
While the standard deviation is very small ($\leq 0.04$) indicating no degradation of the connection for the recorded temperature cycles, the mean reflected power increases significantly (by $14\%$) at $\SI{77}{\kelvin}$.
We attribute the shift purely to thermal contraction effects that can partially be compensated for by using a spring-loaded connector for the fiber applying a constant force to the connection.
Although this change in reflected power cannot be directly converted to an absolute shift as its direction is unknown, we can give a lower bound of $\SI{0.13}{\micro\meter}$ to the shift. 
We expect the absolute shift to be in the order of $\sim \SI{0.2}{\micro\meter}$, as larger shifts would likely cause a larger standard deviation at $\SI{77}{\kelvin}$.
Beyond that, as the glass transition temperature of the epoxy adhesive is well above room temperature, no significant structural changes occur between $\SI{4}{\kelvin}$ and $\SI{77}{\kelvin}$.
Thus, the coupling survives thermal cycling to $\SI{4}{\kelvin}$ without significant loss of performance.
Combining the different shift contributions from alignment precision, concentricity, mating cycles and thermal cycles we can calculate the total expected shift to be $1.4 \pm \SI{0.4}{\micro\meter}$. 

\section{Conclusion}
In summary, we have demonstrated a reliable mechanism for aligning and butt coupling optical fibers to on-chip photonic devices.
Here, a 1D line scan maps out the device center with a precision of $1.2\pm\SI{0.4}{\micro\meter}$, similar to that of DRIE-process based connections\cite{Miller:11}.
With an alignment time $<\SI{1}{\minute}$, the presented mechanism is limited only by the curing time of the used adhesive.
The choice of the adhesive to fix the sample to the ferrule has to be made carefully and situation-dependent. 
Especially cryogenic environments are hereby highly demanding. 
Here, the employed epoxy glue did not show any signs of fatigue with a thermal cycling stability $<\SI{0.2}{\micro\meter}$. 
Moreover, the average shift caused by multiple mating procedures $<\SI{0.5}{\micro\meter}$ is in the order of the precision of the concentricity of the employed fiber. 
These shifts combined exhibit an expected shift of $1.4 \pm \SI{0.4}{\micro\meter}$. 
Its reversibility and compactness make the presented fiber coupling mechanism ideal for the use in environments with limited space such as cryostats.
Its speed, ease-of-use and applicability to all kinds of substrate materials as well as its potential to be fully automated makes it a genuine alternative to other fiber-to-device coupling mechanisms\cite{Bardalen2015}.

\acknowledgments
We gratefully acknowledge the the German Federal Ministry of Education and Research via the funding programs Q.com (contract number 16KIS0110), MARQUAND (contract number BN105022), Photonics Research Germany (contract number 13N14846) and the Deutsche Forschungsgemeinschaft (DFG, German Research Foundation) under Germany’s Excellence Strategy – EXC-2111 – 390814868.

\bibliographystyle{spiebib}
\bibliography{references}

\begin{thebibliography}{10}

\bibitem{Kim08}
Kimble, H., ``The quantum internet,'' {\em Nature}~{\bf 453},  1023--1030
  (2008).

\bibitem{Svo2016}
Svore, K.~M. and Troyer, M., ``The quantum future of computation,'' {\em
  Computer}~{\bf 49}(9),  21--30 (2016).

\bibitem{Sin2016}
Singh, J. and Singh, M., ``Evolution in quantum computing,'' in [{\em 2016
  International Conference System Modeling Advancement in Research Trends
  (SMART)}{\nolinebreak\hspace{0.1em}]},   267--270 (2016).

\bibitem{BB84}
Bennett, C. and Brassard, G., ``Quantum cryptography: Public-key distribution
  and coin tossing,'' {\em Proceedings of IEEE International Conference on
  Computers, Systems and Signal Processing, Bangalore, 175-179} (1984).

\bibitem{Tak2007}
Takesue, H., Nam, S.~W., Zhang, Q., Hadfield, R.~H., Honjo, T., Tamaki, K., and
  Yamamoto, Y., ``{Quantum key distribution over a 40-dB channel loss using
  superconducting single-photon detectors},'' {\em Nature Photonics}~{\bf
  1}(6),  343--348 (2007).

\bibitem{Shi2014}
Shibata, H., Honjo, T., and Shimizu, K., ``Quantum key distribution over a 72
  db channel loss using ultralow dark count superconducting single-photon
  detectors,'' {\em Opt. Lett.}~{\bf 39},  5078--5081 (Sep 2014).

\bibitem{Pfa2014}
Pfaff, W., Hensen, B.~J., Bernien, H., van Dam, S.~B., Blok, M.~S., Taminiau,
  T.~H., Tiggelman, M.~J., Schouten, R.~N., Markham, M., Twitchen, D.~J., and
  Hanson, R., ``Unconditional quantum teleportation between distant solid-state
  quantum bits,'' {\em Science}~{\bf 345}(6196),  532--535 (2014).

\bibitem{Cal2019}
Calvo, R.~M., Poliak, J., Surof, J., Reeves, A., Richerzhagen, M., Kelemu,
  H.~F., Barrios, R., Carrizo, C., Wolf, R., Rein, F., Dochhan, A., Saucke, K.,
  and Luetke, W., ``{Optical technologies for very high throughput satellite
  communications},'' in [{\em Free-Space Laser Communications
  XXXI}{\nolinebreak\hspace{0.1em}]},   {\bf 10910},  189 -- 204, International
  Society for Optics and Photonics, SPIE (2019).

\bibitem{Iva2020}
Ivanov, H.~D. and Leitgeb, E., ``Characteristics of ultra-long deep space fso
  downlinks using special detector technologies like snspd,'' in [{\em 2020
  22nd International Conference on Transparent Optical Networks, ICTON
  2020}{\nolinebreak\hspace{0.1em}]},  {\em International Conference on
  Transparent Optical Networks} (jul 2020).
\newblock 2020 22nd International Conference on Transparent Optical Networks :
  ICTON 2020.

\bibitem{Fry2000}
Fry, P.~W., Finley, J.~J., Wilson, L.~R., Lemaître, A., Mowbray, D.~J.,
  Skolnick, M.~S., Hopkinson, M., Hill, G., and Clark, J.~C.,
  ``Electric-field-dependent carrier capture and escape in self-assembled
  inas/gaas quantum dots,'' {\em Applied Physics Letters}~{\bf 77}(26),
  4344--4346 (2000).

\bibitem{Kur2000}
Kurtsiefer, C., Mayer, S., Zarda, P., and Weinfurter, H., ``Stable solid-state
  source of single photons,'' {\em Phys. Rev. Lett.}~{\bf 85},  290--293 (Jul
  2000).

\bibitem{Tra2016}
Tran, T.~T., Elbadawi, C., Totonjian, D., Lobo, C.~J., Grosso, G., Moon, H.,
  Englund, D.~R., Ford, M.~J., Aharonovich, I., and Toth, M., ``Robust
  multicolor single photon emission from point defects in hexagonal boron
  nitride,'' {\em ACS nano}~{\bf 10},  7331—7338 (August 2016).

\bibitem{Nix1932}
{Nix}, F.~C., ``{Photo-conductivity},'' {\em Reviews of Modern Physics}~{\bf
  4},  723--766 (Oct. 1932).

\bibitem{Ull2015}
Ullom, J.~N. and Bennett, D.~A., ``Review of superconducting transition-edge
  sensors for x-ray and gamma-ray spectroscopy,'' {\em Superconductor Science
  and Technology}~{\bf 28}(8),  084003 (2015).

\bibitem{Gol01}
Gol'tsman, G., Okunev, O., Chulkova, G., Lipatov, A., Dzardanov, A., Smirnov,
  K., Semenov, A., Voronov, B., Williams, C., and Sobolewski, R., ``Fabrication
  and properties of an ultrafast nbn hot-electron single-photon detector,''
  {\em IEEE Transactions on Applied Superconductivity}~{\bf 11}(574-577)
  (2001).

\bibitem{Rei13}
Reithmaier, G., Senf, J., Lichtmannecker, S., Reichert, T., Flassig, F., Voss,
  A., Gross, R., and Finley, J.~J., ``Optimisation of nbn thin films on gaas
  substrates for in-situ single photon detection in structured photonic
  devices,'' {\em Journal of Applied Physics}~{\bf 113}(143507) (2013).

\bibitem{Miller:11}
Miller, A.~J., Lita, A.~E., Calkins, B., Vayshenker, I., Gruber, S.~M., and
  Nam, S.~W., ``Compact cryogenic self-aligning fiber-to-detector coupling with
  losses below one percent,'' {\em Opt. Express}~{\bf 19},  9102--9110 (May
  2011).

\bibitem{Redaelli_2016}
Redaelli, L., Bulgarini, G., Dobrovolskiy, S., Dorenbos, S.~N., Zwiller, V.,
  Monroy, E., and G{\'{e}}rard, J.~M., ``Design of broadband high-efficiency
  superconducting-nanowire single photon detectors,'' {\em Superconductor
  Science and Technology}~{\bf 29},  065016 (may 2016).

\bibitem{Lee2019}
Lee, C.-M., Buyukkaya, M.~A., Aghaeimeibodi, S., Karasahin, A., Richardson, C.
  J.~K., and Waks, E., ``A fiber-integrated nanobeam single photon source
  emitting at telecom wavelengths,'' {\em Applied Physics Letters}~{\bf
  114}(17),  171101 (2019).

\bibitem{Kupko2020}
Kupko, T., von Helversen, M., Rickert, L., Schulze, J.-H., Strittmatter, A.,
  Gschrey, M., Rodt, S., Reitzenstein, S., and Heindel, T., ``Tools for the
  performance optimization of single-photon quantum key distribution,'' {\em
  npj Quantum Information}~{\bf 6},  29 (Mar 2020).

\bibitem{kupko2021evaluating}
Kupko, T., Rickert, L., Urban, F., Große, J., Srocka, N., Rodt, S., Musiał,
  A., Żołnacz, K., Mergo, P., Dybka, K., Urba\'{n}czyk, W., S\c{e}k, G.,
  Burger, S., Reitzenstein, S., and Heindel, T., ``Evaluating a stand-alone
  quantum-dot single-photon source for quantum key distribution at telecom
  wavelengths,'' (2021).

\bibitem{Strauf2010}
Strauf, S., ``Towards efficient quantum sources,'' {\em Nature Photonics}~{\bf
  4},  132--134 (Mar 2010).

\bibitem{Mantynen2019}
Mäntynen, H., Anttu, N., Sun, Z., and Lipsanen, H., ``Single-photon sources
  with quantum dots in iii–v nanowires,'' {\em Nanophotonics}~{\bf 8}(5),
  747--769 (2019).

\bibitem{Comyn2018}
Comyn, J.,  [{\em Thermal Properties of
  Adhesives}{\nolinebreak\hspace{0.1em}]},  459--487, Springer International
  Publishing, Cham (2018).

\bibitem{Mik2013}
Miki, S., Yamashita, T., Terai, H., and Wang, Z., ``High performance
  fiber-coupled nbtin superconducting nanowire single photon detectors with
  gifford-mcmahon cryocooler,'' {\em Opt. Express}~{\bf 21},  10208--10214 (Apr
  2013).

\bibitem{Zha17}
Zhang, W., You, L., Li, H., Huang, J., Lv, C., Zhang, L., Liu, X., Wu, J.,
  Wang, Z., and Xie, X., ``Nbn superconducting nanowire single photon detector
  with efficiency over 90
  cryocooler temperature,'' {\em Science China Physics}~{\bf 60}(120314)
  (2017).

\bibitem{Lie2014}
Liebermeister, L., Petersen, F., Münchow, A.~v., Burchardt, D., Hermelbracht,
  J., Tashima, T., Schell, A.~W., Benson, O., Meinhardt, T., Krueger, A.,
  Stiebeiner, A., Rauschenbeutel, A., Weinfurter, H., and Weber, M., ``Tapered
  fiber coupling of single photons emitted by a deterministically positioned
  single nitrogen vacancy center,'' {\em Applied Physics Letters}~{\bf 104}(3),
   031101 (2014).

\bibitem{Bur2016}
Burek, M.~J., Meuwly, C., Evans, R.~E., Bhaskar, M.~K., Sipahigil, A., Meesala,
  S., Machielse, B., Sukachev, D.~D., Nguyen, C.~T., Pacheco, J.~L., Bielejec,
  E., Lukin, M.~D., and Lon\ifmmode~\check{c}\else \v{c}\fi{}ar, M.,
  ``Fiber-coupled diamond quantum nanophotonic interface,'' {\em Phys. Rev.
  Applied}~{\bf 8},  024026 (Aug 2017).

\bibitem{Tie2015}
Tiecke, T.~G., Nayak, K.~P., Thompson, J.~D., Peyronel, T., de~Leon, N.~P.,
  Vuleti\'{c}, V., and Lukin, M.~D., ``Efficient fiber-optical interface for
  nanophotonic devices,'' {\em Optica}~{\bf 2},  70--75 (Feb 2015).

\bibitem{Mik2010}
Miki, S., Yamashita, T., Fujiwara, M., Sasaki, M., and Wang, Z., ``Multichannel
  snspd system with high detection efficiency at telecommunication
  wavelength,'' {\em Opt. Lett.}~{\bf 35},  2133--2135 (Jul 2010).

\bibitem{Wan2011}
Wang, Q., Loh, T.-H., Ng, D. K.~T., and Ho, S.-T., ``Design and analysis of
  optical coupling between silicon nanophotonic waveguide and standard
  single-mode fiber using an integrated asymmetric super-grin lens,'' {\em IEEE
  Journal of Selected Topics in Quantum Electronics}~{\bf 17}(3),  581--589
  (2011).

\bibitem{Geh2019}
Gehring, H., Blaicher, M., Hartmann, W., Varytis, P., Busch, K., Wegener, M.,
  and Pernice, W. H.~P., ``Low-loss fiber-to-chip couplers with ultrawide
  optical bandwidth,'' {\em APL Photonics}~{\bf 4}(1),  010801 (2019).

\bibitem{Geh2019.2}
Gehring, H., Eich, A., Schuck, C., and Pernice, W. H.~P., ``Broadband
  out-of-plane coupling at visible wavelengths,'' {\em Opt. Lett.}~{\bf 44},
  5089--5092 (Oct 2019).

\bibitem{Xux2021}
Xu, Y., Kuzmin, A., Knehr, E., Blaicher, M., Ilin, K., Dietrich, P.-I., Freude,
  W., Siegel, M., and Koos, C., ``Superconducting nanowire single-photon
  detector with 3d-printed free-form microlenses,'' {\em Opt. Express}~{\bf
  29},  27708--27731 (Aug 2021).

\bibitem{patent:5501893}
Laermer, F. and Schilp, A., ``Method of anisotropically etching silicon,''
  (March 1996).
\newblock Patent DE4241045C1.

\bibitem{Zad2021}
Esmaeil~Zadeh, I., Chang, J., Los, J. W.~N., Gyger, S., Elshaari, A.~W.,
  Steinhauer, S., Dorenbos, S.~N., and Zwiller, V., ``Superconducting nanowire
  single-photon detectors: A perspective on evolution, state-of-the-art, future
  developments, and applications,'' {\em Applied Physics Letters}~{\bf
  118}(19),  190502 (2021).

\bibitem{Thorlabs-Y-Splitter}
``Y-splitter needed for alignment, wavelength dependent..''
  \url{https://www.thorlabs.com/navigation.cfm?guide_id=2421}.
\newblock Accessed: 2022-01-06.

\bibitem{sg}
Savitzky, A. and Golay, M. J.~E., ``Smoothing and differentiation of data by
  simplified least squares procedures.,'' {\em Analytical Chemistry}~{\bf
  36}(8),  1627--1639 (1964).

\bibitem{Pig2014}
Piggott, A.~Y., Lu, J., Babinec, T.~M., Lagoudakis, K.~G., Petykiewicz, J., and
  Vučković, J., ``Inverse design and implementation of a wavelength
  demultiplexing grating coupler,'' {\em Scientific Reports}~{\bf 4} (Nov
  2014).

\bibitem{Fiber780HP}
``Information from thorlabs about the 780hp fiber.''
  \url{https://www.thorlabs.com/drawings/fd94e32bbba7dc0c-D4715921-FB12-4FDE-FD9C7582EBE3468D/780HP-SpecSheet.pdf}.
\newblock Accessed: 2022-01-06.

\bibitem{Bardalen2015}
Bardalen, E., Akram, M.~N., Malmbekk, H., and Ohlckers, P., ``{Review of
  Devices, Packaging, and Materials for Cryogenic Optoelectronics},'' {\em
  Journal of Microelectronics and Electronic Packaging}~{\bf 12},  189--204 (10
  2015).

\end{thebibliography}
\end{document}